\documentclass[%
 reprint,
 amsmath,amssymb,
 aps,letterpaper,superscriptaddress
]{revtex4-2}

\usepackage{graphicx}
\usepackage{dcolumn}
\usepackage{bm}

\usepackage{xcolor}
\newcommand{\change}[1]{{\color{red}#1}}
\renewcommand{\change}[1]{#1}

\usepackage{booktabs}
\usepackage{multirow}
\usepackage{hyperref}
\begin{document}

\preprint{APS/123-QED}

\title{Tuning Colloidal Reactions}

\author{Ryan Krueger}
 \thanks{These authors contributed equally to this work.}
\affiliation{%
School of Engineering and Applied Sciences, Harvard University,
29 Oxford St, Cambridge, MA 02138, USA
}%

\author{Ella King}
\thanks{These authors contributed equally to this work.}
\affiliation{
 Department of Physics, New York University,
 726 Broadway, New York, NY 10003, USA
}%
\affiliation{
 Simons Junior Fellow,
 160 5th Ave, New York, NY 10010, USA
}
\author{Michael Brenner}
\email{brenner@seas.harvard.edu}
\affiliation{%
School of Engineering and Applied Sciences, Harvard University,
29 Oxford St, Cambridge, MA 02138, USA
}%


\date{\today}

\begin{abstract}
The precise control of complex reactions is critical for biological processes yet our inability to \change{design for specific} outcomes limits the development of synthetic analogues.
Here, we leverage differentiable simulators to design nontrivial reaction pathways in colloidal assemblies.
By optimizing \change{over external structures}, we achieve controlled disassembly \change{and particle release from} \change{colloidal} shells.
\change{
Lastly, we characterize the role of configurational entropy in the structure via both forward calculations and optimization,
inspiring new parameterizations of designed colloidal reactions.
}
\end{abstract}

\maketitle

Both living and non-living physical systems exhibit complex dynamical behavior, ranging from repair to locomotion to catalysis.
Fundamentally, such behaviors arise from sequences of reactions, in which a set of substances (i.e. the reactants) are transformed into a set of different substances (i.e. the products).
\change{A rich body of work aims to characterize and tune systems of interacting agents spanning a range of system descriptions, both theoretically~\cite{van2015digital, zeravcic2017colloquium, sherman2020inverse} and experimentally~\cite{truong2022light, li2019light, wang2012colloids, mcmullen2022self, niu2019magnetic, schade2013tetrahedral}.
However, for many critical processes in biological systems (e.g. DNA synthesis, protein folding), the components themselves cannot be changed.
Instead, to modify these processes, researchers often introduce an external structure (e.g. competitive inhibitors for enzyme inhibition, protein folding chaperones).
Thus far, the design of \change{such} structures has been bespoke and application-specific, necessitating entirely new research programs for each new reaction.
\change{For example,} while some general theoretical models have provided deep insights into catalysis~\cite{zeravcic2017spontaneous, munoz2023computational}, they are largely too abstract to inform experimental design.}

To \change{overcome current limitations and} tune reactions \change{through the design of external agents, we} carry out {\sl inverse design} whereby we
\change{optimize the geometry and interactions of such components}
to achieve a target reaction.
While inverse design has been successfully applied to self-assembly \cite{torquato2009inverse,chen2018inverse,rechtsman2006designed,ma2019inverse,jaxmd_patchy}, inverse-designing reaction pathways
\change{remains a challenge}
because design parameters must be chosen to favor particular dynamical trajectories.
The advent of differentiable simulators \cite{jaxmd_patchy}, powered by software libraries developed for machine learning \cite{jax2018github}, has opened up the possibility of directly designing reactions
\change{as the gradient of numerical procedures with respect to control parameters can be computed efficiently.}

\change{Here, we design complex reactions using differentiable molecular dynamics (MD) and gradient-based optimization.}
\change{As an example of a nontrivial reaction}, we consider the controlled disassembly of colloidal structures, whereby a particle is extracted from an otherwise complete shell of colloidal particles.
Disassembly is central to the dynamic functions of living systems, such as defect repair, self-replication, and catalysis.
Existing examples of controlled disassembly in synthetic systems often rely on external forcing to drive the disassembly process~\cite{jung2023magnetic, tottori2013assembly, martinez2020controlled, kostiainen2011hierarchical}, which provides a direct pathway to tuning behavior.
However, for many engineering applications including those inherent to living systems, the use of external fields is limiting.
On the other hand, controlled disassembly in living systems typically relies on local energy consumption (e.g. biological enzymes consuming ATP) rather than global fields, but the synthetic design of these systems is significantly more complex.

\change{Inspired by the symmetry of many viral capsids~\cite{caspar1962physical, twarock2019structural}}, we design for the controlled disassembly
\change{of icosahedral shells}.
\change{We consider a fixed shell and only parameterize an external structure that acts upon it, enabling control over disassembly without modifying critical components of the reaction.
Importantly,} our disassembly mechanism \change{is entirely passive and} does not rely on external forcing.
As a model for potential engineering applications, we apply our mechanism to provoke the release of a target small particle \change{initially} trapped inside \change{the} shell.
Controlled disassembly serves here as a striking example of a complex reaction because the reaction requires a finely-tuned interaction energy to keep the remaining shell stabilized while still performing the desired particle extraction.
\change{We start from a rigid structure and thereafter proceed to quantify the role of flexibility in the structure by computing free energy landscapes both for predefined extrema and for structures optimized via a chosen parameterization of configurational entropy, opening the door to novel designed reactions.}

\begin{figure}[t]
\begin{center}
\centerline{\includegraphics[width=1.0\columnwidth]{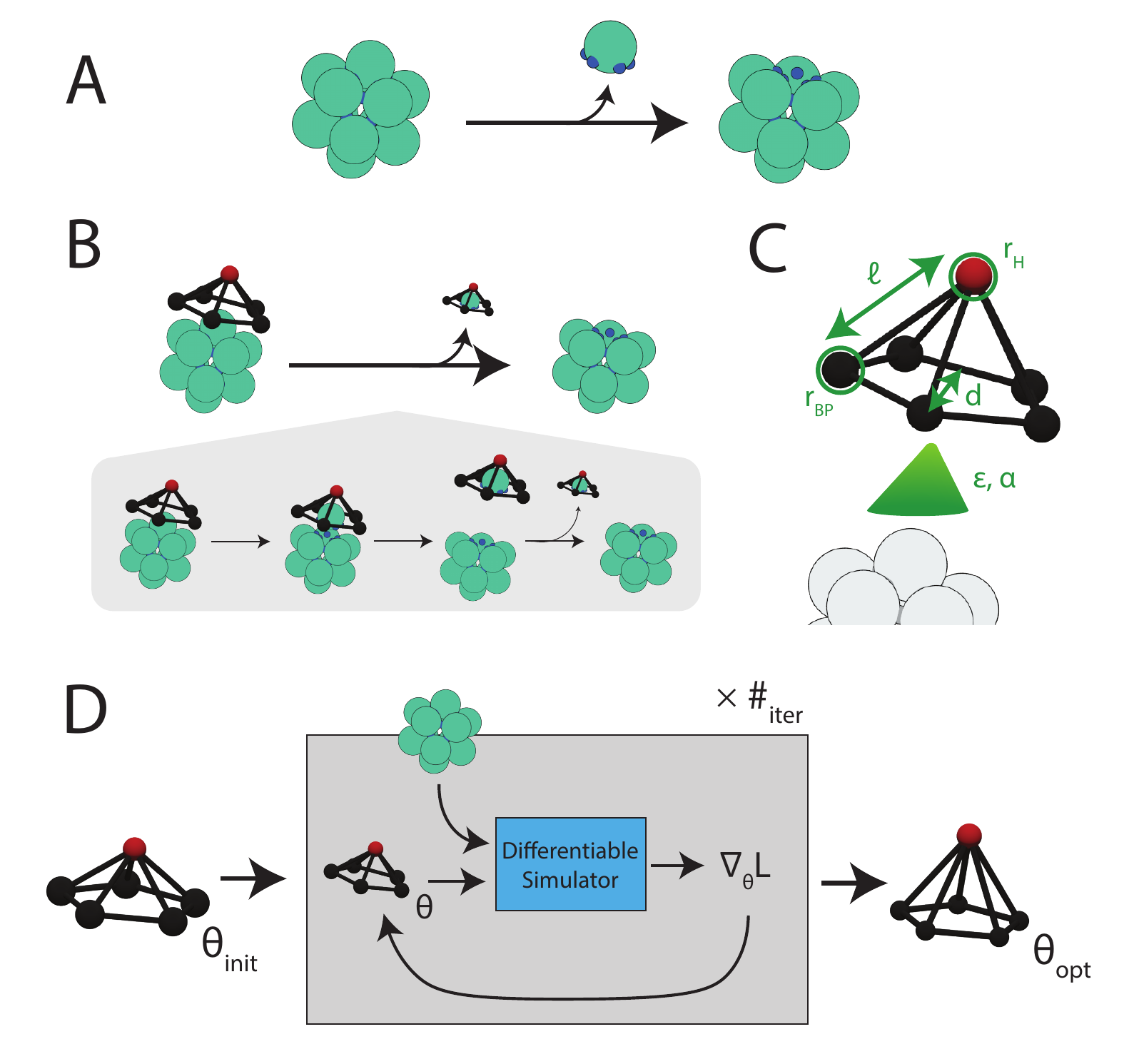}}
\caption{
\textbf{Tuning the interaction potential of an external structure, the spider, to achieve a desired reaction of disassembly.}
\textbf{A.} A single particle is removed from an \change{icosahedron}.
\textbf{B.} A candidate mechanism: the spider extracts the target particle via an attractive potential and detaches from the remaining shell.
\textbf{C.} Parameterization of the spider geometry and interaction potential with the shell.
The red particle is the ``head'' particle, situated above the four black ``base'' particles that constitute the ring.
The interaction energy between spider and shell is depicted as a green triangle.
We optimize over all labeled parameters, as well as the cutoff of the interaction energy (not depicted).
\textbf{D.} High-level depiction of our optimization pipeline: analytic gradients are computed via a differentiable molecular dynamics simulator and parameters of the spider are updated via gradient descent.
}
\label{fig:overview}
\end{center}
\vspace{-25pt}
\end{figure}

\begin{figure*}[t]
\centering
\includegraphics[width=1.0\textwidth]{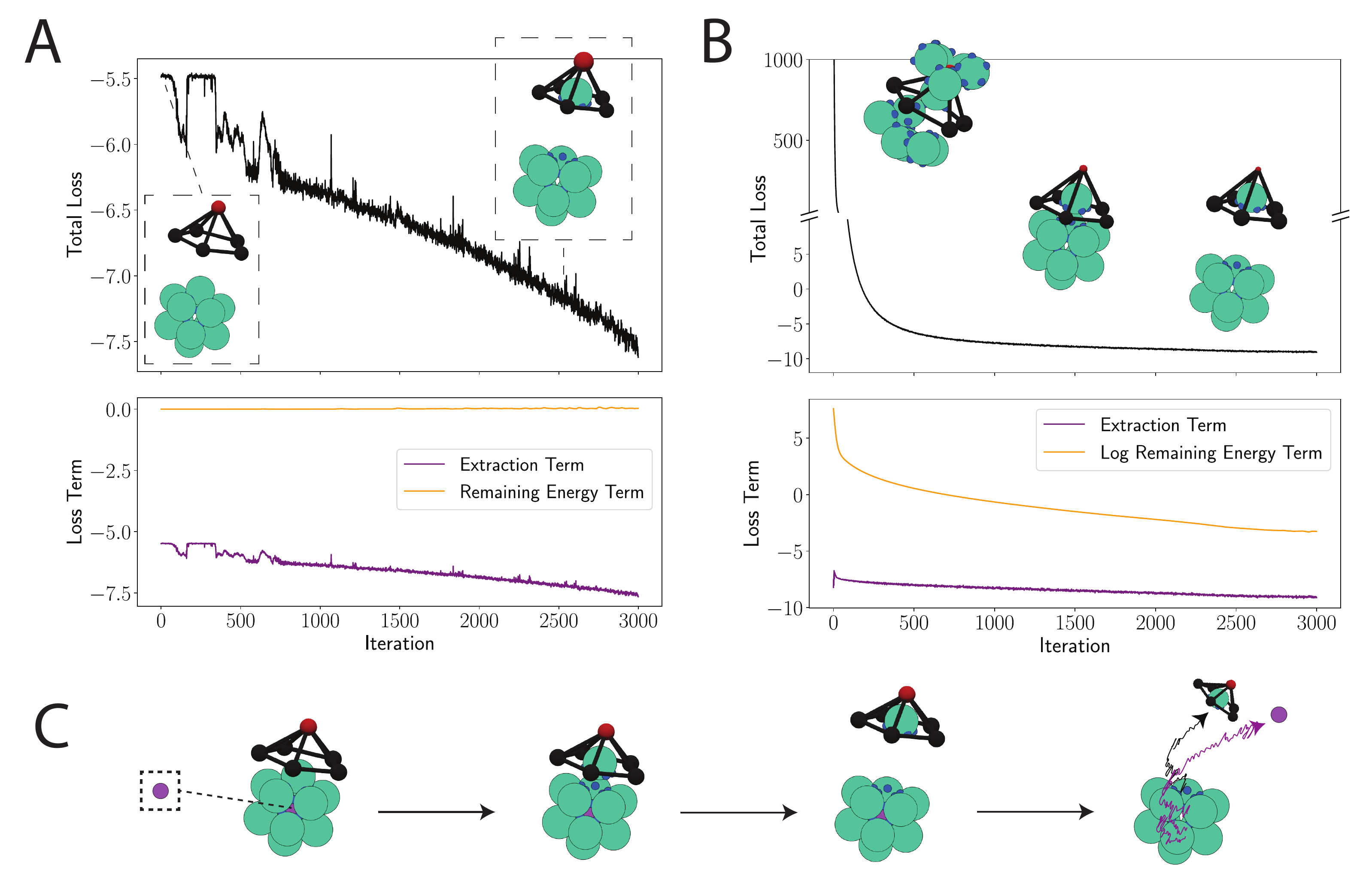}
\vspace{-5pt}
\caption{
\textbf{\change{Optimizing the geometry and interaction potential of a rigid spider}.}
\textbf{A.} In the limit of a weak initial spider-shell interaction, the initial spider simply diffuses away from the shell.
By the \change{3000th} iteration, the spider geometry and interaction energy are optimized to extract the target particle while still diffusing away from the remaining particles, leaving them intact.
\textbf{B.} In the limit of a strong initial spider-shell interaction, the initial spider extracts the target particle but does not diffuse away, disturbing the integrity of the remaining shell.
As optimization progresses, the interaction is tuned to only extract the particle without disrupting the remaining shell.
Upon convergence, the spider geometry and interaction energy are tuned to maintain extraction while diffusing away.
Insets depict representative states after 10,000 MD steps at the corresponding iteration.
\change{\textbf{C.} Schematic of a simulation of an optimized spider provoking the release of a particle from an icosahedral shell.}}
\label{fig:rigidspider}
\end{figure*}


\textit{Results.---}We implement controlled disassembly in a \change{colloidal} patchy particle system.
Patchy particles have long been used to emulate interactions in soft materials~\cite{zhang2004self,glotzer2007anisotropy} and offer tremendous \change{tunability} in designed interactions.
Optimizing said systems to achieve specific behaviors has been made possible  by the recent development of patchy particle simulations within a differentiable library \cite{jaxmd_patchy}.
In particular, we aim to remove a single particle from a shell composed of patchy particles in a controlled manner without disrupting the remaining shell structure.
\change{To that end,} we tune disassembly without changing any properties of the shell itself.
Instead, we introduce an external \change{structure} that interacts with the shell to disassemble it in the desired manner.
We term the external structure a ``spider'' \change{due to its geometry}.

The shell is modelled as a collection of patchy particles forming an icosahedron where each patch corresponds to a contact with a neighboring particle.
Each \change{patchy particle} consists of a \change{central sphere} and a set of rigidly attached patches. Patches interact via a Morse potential ($\epsilon_{V} = 10.0$, $\alpha_{V} = 5.0$) and particles interact via soft-sphere repulsion ($\epsilon_{ss} = 10^4$).
Importantly, the geometry and interaction energy of shell-comprising particles are fixed throughout the optimization. 
\change{Although we focus on the disassembly of the icosahedron, our framework can be easily adapted for octahedral shells (see Appendix).}

\change{
For the spider, we consider several different models with a set of core similarities.
All models contain a ring of ``base'' particles and a ``head'' particle that sits above the ring along its symmetry axis.
}
The head is connected to the base particles by repulsive bars, making the entire structure a cage-like object that is open on one end.
\change{An attractive particle type (either the head or a third particle species)}
interacts with the shell-comprising particles via a Morse potential whereas base particles and connecting bars interact with shell particles via soft sphere repulsion.
Unlike the shell, the geometry and interaction energy of the spider are parameters of the optimization.
See Figure \ref{fig:overview}C for an overview of this parameterization.
All interaction energies in our system are parameterized with simple, physics-based potentials.

\change{Given a specified parameterization for the spider geometry and interactions, we run an ensemble of differentiable molecular dynamics simulations (see Figure \ref{fig:overview}D and Appendix).
To focus our optimization procedure on the challenges specific to disassembly, we initialize the spider bound to the shell and therefore ignore the period in which the spider is freely diffusing.}
We optimize over \change{the} parameters that characterize the geometry of the spider and its interaction with the shell \change{(8 parameters for the optimizations in Figure \ref{fig:rigidspider})}.
To optimize our system, we perform gradient descent to minimize a loss function.
The loss is constructed from two competing terms: one that rewards a final state in which the target particle is extracted, and one that penalizes a strong interaction between the spider and non-target particles.
The second term, which we refer to as the ``remaining energy'' term,  tends to reward pathways in which the remaining shell holds its shape after the spider extracts the target particle.

We formalize the loss function as follows.
Consider an icosahedral shell comprised of a collection of particles $V = \{\overrightarrow{v}_1, \overrightarrow{v}_2, \cdots, \overrightarrow{v}_n\}$ where $n=12$.
We seek to extract a target particle $\overrightarrow{v}_j$ from the shell while leaving the remaining shell $V \setminus \overrightarrow{v}_j$ intact.
We can measure the degree to which $\overrightarrow{v}_j$ is successfully extracted via the following expression:
\begin{align}
    \mathcal{L}_{\text{extract}}(V) = -\sum_{i \neq j} d(\overrightarrow{v}_i, \overrightarrow{v}_j)
\end{align}
where $d(\overrightarrow{v}_i, \overrightarrow{v}_j)$ denotes the Euclidean distance between particles $\overrightarrow{v}_i$ and $\overrightarrow{v}_j$.
Note the negative sign as we formulate our optimization problem to minimize the loss.
Next, we minimize the interaction energy between
\change{attractive site(s)}
and non-target shell particles:
\begin{align}
    \mathcal{L}_{\text{remain}}(V, A) = \sum_{\overrightarrow{a} \in A} \left(\sum_{i \neq j} U_{\text{m}}(\overrightarrow{a}, \overrightarrow{v}_i) \right)^2
\end{align}
\change{where $A$ denotes the set of attractive sites and
$U_{\text{m}}(\overrightarrow{a}, \overrightarrow{v}_i)$ represents the interaction energy between the attractive site $\overrightarrow{a}$ and a shell particle $\overrightarrow{v}_i \in V$.
For the spider depicted in Figure \ref{fig:overview}, the head particle is the only attractive site.}
\change{
We calculate the `remaining energy' term, $\mathcal{L}_{\text{remain}}$, of the total loss
with respect to the initial configuration, i.e. the first timestep of the simulation.
All other terms depend on the dynamics of the system, so we evaluate them on the final state.
}
In all simulations, the spider is initially bound to the target particle and we integrate the system for 1000 timesteps (see Appendix).

\begin{figure*}[t]
\begin{center}
\centerline{\includegraphics[width=1.0\textwidth]{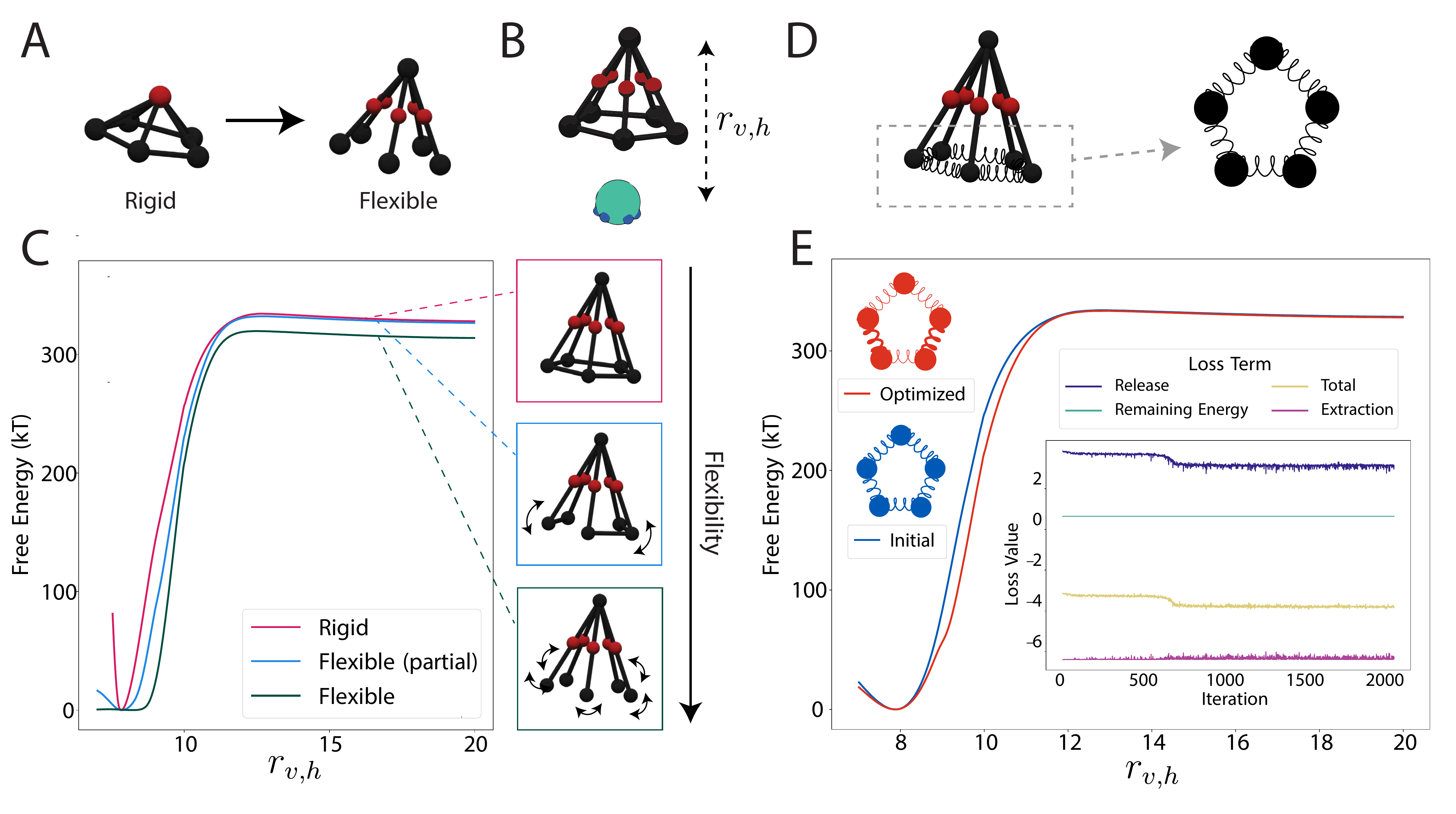}}
\vspace{-10pt}
\caption{
\change{\textbf{Role of configurational entropy in the release of a target particle.}
\textbf{A.} A modified version of the original spider in which a ring of attractive sites are positioned between the base and head particles on each spider leg.
\textbf{B.} We use the distance from the extracted particle to the spider head as the order parameter approximating particle release.
\textbf{C.} Free energy diagrams for the order parameter depicted in (B) for three variants of the modified spider: (i) fully rigid, (ii) partially flexible, defining two rigid substructures which are free to rotate about the head, and (iii) fully flexible, in which all base bonds are removed.
The parameters for the spider geometry and interaction are determined via an optimization of the rigid configuration.
Free energy diagrams are computed via the Weighted Histogram Analysis Method (WHAM).
\textbf{D.} A parameterization of the configurational entropy of the modified spider in which the base bonds are represented as springs whose spring constants are free parameters.
Identical springs are also placed between all next-nearest neighbors to parameterize the bond angles (not shown).
\textbf{E.} The free energies of the initial spider with uniform spring constant and the optimized spider as computed via WHAM. Inset: an optimization over the spring constants depicted in (D) to maintain extraction and minimize remaining energy while also minimizing the average energy between extracted particle and attractive sites over an additional 500 simulation steps.
The parameters defining the spider geometry and interaction are set to those used to compute free energies in (C).
}
}
\label{fig:flexible}
\end{center}
\vspace{-20pt}
\end{figure*}

\change{\textit{Rigid Spider.--}}
\change{
We begin with a minimal conception of the spider: a rigid body consisting of only a head particle and five base particles which reflect the five-fold symmetry of the icosahedron.
The base particles are attached to each other by rigid bars, forming a cage-like structure that is open on one end.
The attractive interactions between the spider and the icosahedron are restricted to interactions between the head particle in the spider and the patches on the icosahedral vertices.
}

We explore two limits of our optimization procedure (Figure \ref{fig:rigidspider}).
First, we perform an optimization where the spider is initialized to interact weakly with the shell particles ($log(\epsilon_{H}) = 3.0$, $\alpha_{H} = 1.5$).
\change{In this limit}, the spider simply diffuses away from the shell at long timescales without extracting the target particle.
\change{Initially, we observe variable changes consistent with increasing the interaction between spider head and shell to achieve particle extraction: $\epsilon_{H}$ increases, the head height decreases, and the head particle radius increases.}
\change{
In the following iterations, we observe parameter changes focused on maintaining extraction while reducing the interaction strength between the spider and the rest of the shell.
The head height increases, consistent with minimizing the remaining energy, but to maintain particle extraction, $\alpha_H$ decreases (increasing the range of the Morse potential).}
This suggests that tightly coupled, nontrivial parameter changes drive extraction while maintaining minimal interaction with the remaining shell.

Next, we perform an optimization in the opposite limit in which the spider is initialized to interact strongly with the shell ($log(\epsilon_{H}) = 10.5$, $\alpha_{H} = 1.5$).
Initially, this interaction is so strong that the spider not only extracts the target particle but it also disrupts the remaining shell.
This can be seen in the large value of the remaining energy loss term, which penalizes the energy between the spider head and non-target particles.
Throughout the optimization, we observe variable changes consistent with tuning the interaction strength to maintain extraction while minimizing off-target interactions: $\epsilon_{H}$ decreases, $\alpha_{H}$ increases, the head radius decreases, the head height increases, and the base particle radius increases.
When evaluated on longer simulations,
the converged parameter set also achieves spontaneous diffusion of the spider-particle complex from the remaining undisturbed shell.
\change{Note that neither changes in the random seed nor perturbations to the initial parameters consistently yield similar optimized parameters (see Appendix).}

Contrasting the high and low energy optimization regimes reveals the inherent delicacy in tuning the spider to achieve extraction and subsequent diffusion from the shell.
The spider-shell interaction must be sufficiently strong to extract the target particle, but simultaneously weak enough to not disturb non-target particles and to diffuse away from the shell within the timescale of our simulations.
This tension is reflected in the behavior of the loss terms in each optimization.
Overall, in the weak-interaction limit, the term penalizing interactions with non-target particles remains negligible while the extraction term drives optimization; in the strong-interaction limit it is the same energy-penalizing term that dominates the loss.
\change{Our optimized reactions represent a notion of balance that is necessary for biologically relevant functions, e.g. the controlled release of a particle from a closed shell (see simulation in Figure \ref{fig:rigidspider}C).}
This serves as a toy example of a potential target for engineering applications, such as drug delivery via a viral shell.

\change{\textit{Flexible Spider.--}
The configurational entropy of the spider can serve as a control knob for tuning reactions.
While the optimized results in Figure \ref{fig:rigidspider} highlight that spider geometry dramatically impacts its performance, the rigid formulation does not access configurational entropy.
Here, we define a modified form of our spider permitting versions with varying degrees of flexibility.
Rather than the head serving as the sole attractive site, we introduce a ring of attractive sites consisting of one site per spider leg positioned between the base and head particles (Figure \ref{fig:flexible}A).
In this way, bonds at the spider base connecting individual legs can be made flexible or removed entirely.
The resulting fluctuations due to leg flexibility directly change the interaction strength between spider and extracted particle and thus modulate the entropic contribution to the interaction.
In this scheme, the head particle only interacts repulsively with the icosahedron.

Instead of considering the probability of extraction, we focus on the release of an already extracted particle since this process is likely to be strongly influenced by configurational entropy.
We reason that increased entropy in the extracted state (i.e. the extracted particle bound to the spider) would favor particle release compared to the fully rigid spider because fluctuations in the spider configuration would reduce the effective attraction.
To test this hypothesis, we quantitatively measure free energy differences corresponding to particle release between three versions of the modified spider with varying flexibility: (i) a fully rigid spider with fixed bonds between all adjacent base particles, (ii) a partially flexible spider resulting from the removal of two base bonds, and (iii) a fully flexible spider via the removal of all base bonds.

We compute the free energy of release using each of the three models.
The distance from the extracted particle to the head directly relates to its release.
We therefore use this metric as the order parameter for free energy calculations (Figure \ref{fig:flexible}B).
We compute the free energy diagrams for each spider using the Weighted Histogram Analysis Method (WHAM) and use a fixed set of parameters for the spider geometry and interaction determined via a single optimization with the fully rigid spider (see Appendix).
As expected, the more flexible the geometry, the more favorable the released state:
there is a smaller change in free energy between the attached and released states for more flexible geometries (Figure \ref{fig:flexible}C).

Next, we optimize configurational entropy directly.
We define a spider in which all pairs of (i) adjacent and (ii) next-nearest neighbor base particles are connected with springs whose spring constants are free parameters (Figure \ref{fig:flexible}D).
To bias the optimization procedure towards a spider with an increased likelihood of release, we define an additional loss term representing the total interaction energy between attractive sites and extracted particle:
\begin{align}
    \mathcal{L}_{\text{release}}(\overrightarrow{v}, A) = \sum_{\overrightarrow{a} \in A} \left( U_{\text{m}}(\overrightarrow{a}, \overrightarrow{v}) \right)^2.
\end{align}
The weaker the interaction, the easier the release.
Here, we optimize over a longer (1500 step) simulation than in the rigid case.
We average the new loss term over states sampled from the final 500 simulation steps, while maintaining extraction within the first 1000 steps (see Appendix).
To give the optimization algorithm more freedom to promote release, we rescale the loss describing extraction such that it changes minimally beyond a specified maximum value.
The optimization algorithm can then reduce extraction efficiency without penalty.

We optimize these spring constants using the modified loss function while keeping all other parameters fixed.
We initialize all spring constants to be the same value (i.e. $\log (k) = 2.0$) and fix the spider geometry and interaction parameters to those used in Figure \ref{fig:flexible}C.
The optimized solution has a wider well but maintains the same well depth.
As a result, the free energy difference between the extracted and released states is lower in the optimized configuration than in the initial one at intermediate distances.
Interestingly, the optimization procedure naturally converges to a solution with an asymmetric distribution of spring constants.
Directly tuning this asymmetry is a promising avenue for future work.}

\textit{Discussion.---}
In this Letter, we \change{achieve} nontrivial reactions via \change{designed} \change{external structures}.
We consider the case of controlled disassembly \change{of an icosahedral shell composed of patchy particles}, in which there is an inherent tension between initiating disassembly and maintaining the integrity of the remaining substructure.
\change{
We show how the parameters governing a rigid external structure can be finely tuned to minimize a loss function representing this tension.
We find that the optimized spider provokes particle release.
We then add configurational entropy by introducing a flexible spider geometry, and quantify the influence of flexibility by comparing free energy landscapes for varying degrees of flexibility.
Our framework naturally accommodates parameterizations of configurational entropy.
Upon optimization, a spider with asymmetrically flexible base legs favors release over the initialized uniform configuration.}

Since we optimize directly with respect to the numerically integrated dynamics, our method is general enough to study a wide range of systems.
Foremost, it may enable experimental realizations of theoretical models that were otherwise limited by an inability to finely tune interaction energies.
For example, Ref.~\cite{zeravcic2014self} introduces a model of self-replicating colloidal clusters in which kinetic traps can be avoided by tuning the interaction energies, but dissociation of a new cluster from its parent (a necessary step for replication) required an artificial trigger event in numerical simulations.
In contrast, our designed parameters lead to spontaneous dissociation of the spider-particle complex away from the remaining shell.
The computational flexibility of the method could also easily enable users to restrict the parameter regime to experimentally realizable interactions.
This could be done for DNA coated colloids, e.g., by optimizing the DNA sequences that define the interaction strength~\cite{rogers2020mean, rogers2011direct, matthies2023differentiable}.

While optimizing for such types of reactions, numerical instabilities may arise.
The primary limitation we observe is that gradients become unstable and very large for long simulations.
There are several possible approaches to reducing instability in gradients in such cases.
One standard method to mitigate instabilities in the context of differentiable programming is gradient clipping~\cite{greener2023differentiable, engel2022optimal}.
One could also decrease the total number of timesteps by training an emulator to resolve the dynamics with a larger timestep than is otherwise possible with standard integrators, following similar work for deterministic systems~\cite{sanchez2020learning, allen2022physical, bar2019learning}.
An alternative approach would be to integrate differentiable simulations with enhanced sampling methods to sample low probability events without the need for long simulation times.

We rely on gradient-based optimization due to its scalability and performance.
\change{Naturally, our method scales to larger and more complex systems since (i) gradient calculation via automatic differentiation only requires a single simulation, (ii) reverse-mode scales efficiently with the number of parameters~\cite{baydin2018automatic}, and (iii) the gradient explicitly captures interdependencies which is essential to efficiently tuning complex behavior.}
\change{We envision that our approach and proposed design rules} will be applicable to physical reactions beyond the colloidal regime.

\begin{acknowledgments}
We thank Sam Schoenholz for his work developing JAX-MD and the members of the Brenner Group for helpful discussions.
This work is supported by the Office of Naval Research (ONR N00014-17-1-3029, ONR N00014-23-1-2654),  the NSF Grant DMR-1921619, the NSF AI Institute of Dynamic Systems (\#2112085), and a grant from the Simons Foundation (\#1141499). E.M.K. acknowledges support from a Simons Foundation Junior
Fellowship under Grant Number 1141499.
\end{acknowledgments}

\bibliography{main}

\newpage
\appendix

\end{document}


\preprint{APS/123-QED}

\title{Appendix for ``Tuning Colloidal Reactions''}

\author{Ryan Krueger}
 \thanks{These authors contributed equally to this work.}
\affiliation{%
School of Engineering and Applied Sciences, Harvard University,
29 Oxford St, Cambridge, MA 02138, USA
}%

\author{Ella King}
\thanks{These authors contributed equally to this work.}
\affiliation{
 Department of Physics, New York University,
 726 Broadway, New York, NY 10003, USA
}%
\affiliation{
 Simons Junior Fellow,
 160 5th Ave, New York, NY 10010, USA
}
\author{Michael Brenner}
\email{brenner@seas.harvard.edu}
\affiliation{%
School of Engineering and Applied Sciences, Harvard University,
29 Oxford St, Cambridge, MA 02138, USA
}%


\date{\today}

\maketitle


\noindent The code used to generate all results presented in this manuscript is available at the following GitHub repository: {\color{blue}\url{https://github.com/rkruegs123/tuning-colloidal-reactions}} \\


\textit{Appendix A: Differentiable MD.---}In traditional molecular dynamics simulations, a system of $n$ interacting bodies, typically represented by a vector $\overrightarrow{x} \in \mathcal{R}^{6n}$ describing their positions and momenta, is iteratively propagated through time via a transition function:
\begin{align*}
    \overrightarrow{x}_{t+1} = f(\overrightarrow{x}_t, \theta)
\end{align*}
where $f$ depends on the energy function and numerical integration scheme and $\theta$ are control variables.
Thus, for some fixed time length $N$, a molecular dynamics trajectory can be considered the result of a single numerical calculation.
When written in an automatic differentiation framework, gradients of this calculation can be computed efficiently with respect to $\theta$.
More generally, one can express a continuous and differentiable loss function of a molecular dynamics trajectory, $\mathcal{L}(X)$ where $X = \{\overrightarrow{x_0}, \overrightarrow{x_1}, \cdots, \overrightarrow{x_N}\}$ is a trajectory, and efficiently compute $\nabla_{\theta}\mathcal{L}$.
Here, we use differentiable MD to optimize for the parameters of an external entity to remove one component of a self-assembled structure in such a manner that maintains the integrity of the remaining structure.
More specifically, we use JAX-MD~\cite{schoenholz2020jax}, a general-purpose differentiable MD framework implemented in JAX~\cite{jax2018github}.\\

\textit{Appendix B: The System.---}Each simulation consists of two individual components -- the shell and the external structure (the ``spider'').
Since these components interact, a complete description of the simulated system requires a description of the energy and geometry of (i) the shell, (ii) the spider, and (iii) the shell-spider complex.

\textit{Shell.---}The octahedral and icosahedral shells consist of 8 and 12 anisotropic patchy particles, respectively.
We represent each patchy particle as a rigid body consisting of a central particle and a symmetric ring of attractive patches on one face of that central particle (4 and 5 patches for the octahedron and icosahedron, respectively).
The central particle has radius $r_{V} = 2.0$.
All patches interact via a Morse potential with depth $\varepsilon_{V}=10.0$, width $\alpha_{V}=5.0$, and equilibrium distance $\sigma_{V} = 0.0$.
Central particles repel each other via soft sphere repulsion with $\varepsilon_{ss} = 10^4$.
The shell is always initiated in its self-assembled ground state.
For each shell, we obtain its ground state via energy minimization seeded with an initialized shell whose patchy particles are placed on the shell and point inward, but still must orient to align pairs of attractive patches.
Importantly, the geometry and energy of the entire shell is fixed throughout the optimization procedure.
\begin{figure}[b]
\begin{center}
\vspace{-20pt}
\centerline{\includegraphics[width=1.0\columnwidth]{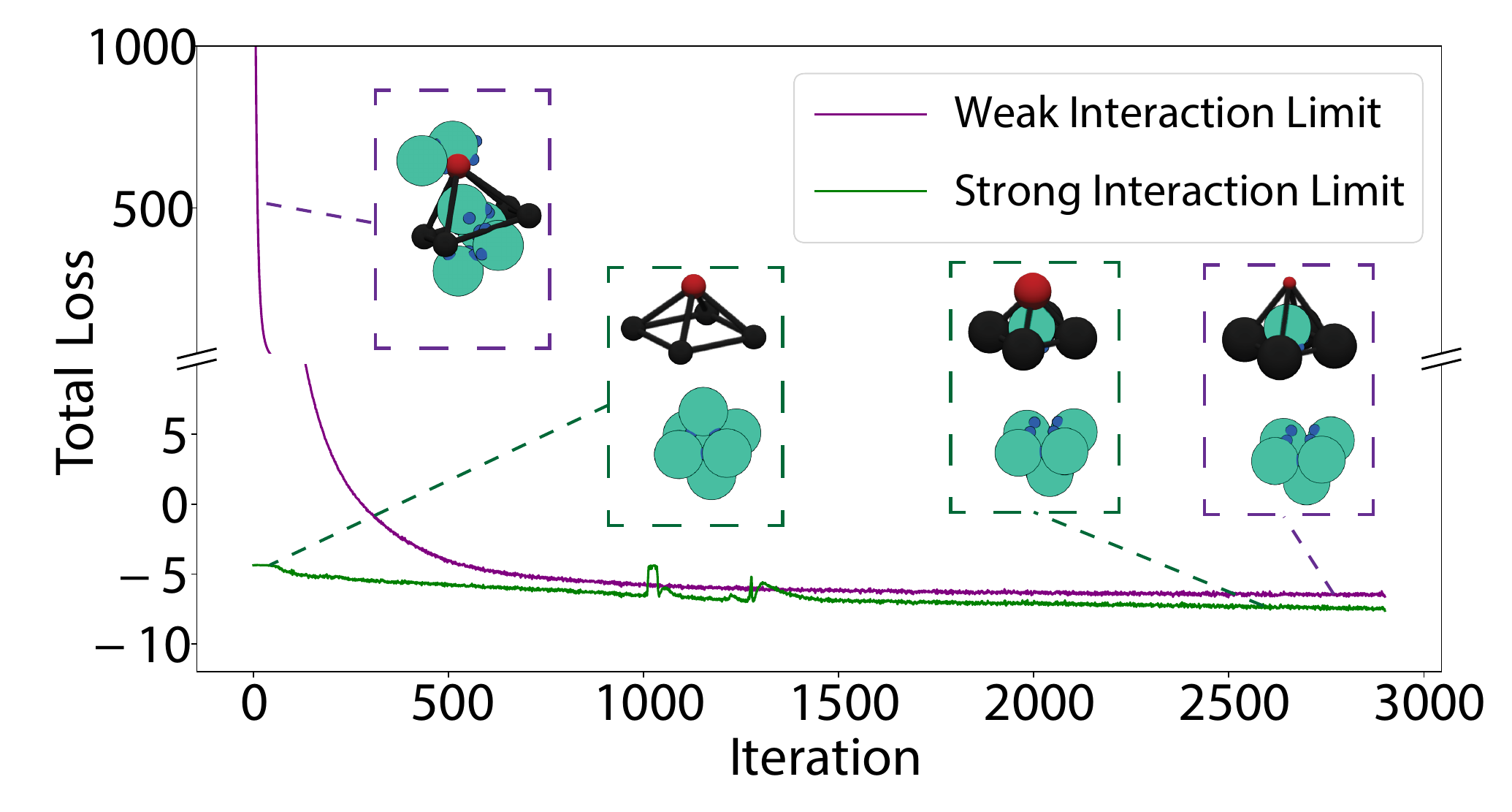}}
\vspace{-10pt}
\caption{
Optimizations for the octahedron in both the weak and strong interaction limits for parameter initialization.
}
\label{fig:octahedron}
\end{center}
\end{figure}

\textit{Spider.---}In the simplest case, we represent the spider as a single rigid body comprised of one head particle connected to several base particles;
the number of base particles depends on the symmetry of the shell to which it is bound (five base particles for the icosahedron, four for the octahedron).
Base particles are connected to each other in sequence (i.e. along the perimeter of the spider base) and to the head via bonds represented by cylinders with radius $r_L = 0.25$. 
We refer to bonds connecting base particles and head particle as ``legs''.
\begin{figure*}[t]
\begin{center}
\centerline{\includegraphics[width=1.0\textwidth]{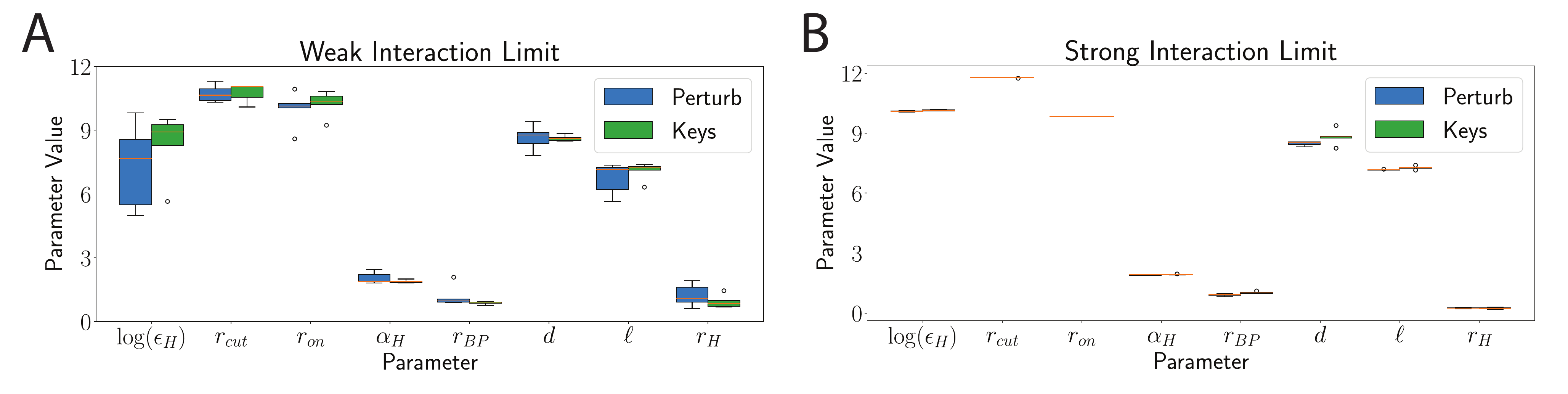}}
\caption{
Evaluating the variation in the optimized parameters for each optimization regime for the icosahedral shell.
For both the \textbf{(A)} weak interaction and \textbf{(B)} strong interaction initial parameter regimes, two sets of five optimizations were performed either with different random keys or with perturbations to the initial value of $\log(\epsilon_H)$ drawn from $\mathcal{N}(0, 0.1)$.
Mean values are depicted by orange lines; batches of optimizations with negligible variance are represented only as orange lines.
}
\label{fig:var-params}
\vspace{-30pt}
\end{center}
\end{figure*}
The geometry of the spider is parameterized by the following variables: (i) the radius of the head particle, $r_{HP}$, (ii) the radius of the base particles, $r_{BP}$, (iii) the diameter of the spider base, $d$, and (iv) the leg length, $\ell$.
Note that fixing $d$ and $\ell$ uniquely determines the distance from the head particle to the spider base (i.e. the head height).
As the spider is a single rigid body, it has no internal energy.

In the modified spider, we also place an additional particle on each leg between the base and head particles.
This introduces two additional parameters for the spider geometry: (1) the radius of these attractive particles, $r_A$, and their relative position on the leg, $p_A \in [0, 1]$.
The modified spider is made flexible by removing selected bonds between base particles.
To simulate this flexible spider, we represent the legs as individual rigid bodies whose head particles interact via a Morse potential with $\epsilon = 10^4$, $\alpha = 4.0$, and $\sigma = 0.0$.

To optimize flexibility, we introduce an alternative version of the modified spider by representing the bonds between base particles with flexible harmonic springs rather than rigid bonds.
Each spring is parameterized via a spring constant $k$ and has equilibrium distance equal to the inter-particle distance in the rigid case.
We also place equivalent springs between all pairs of next-nearest neighbors to parameterize the angle between base bonds.

\textit{Shell-Spider Complex.---}All optimizations are initialized with the spider bound to the shell.
The exact geometry of this shell-spider complex is determined by an initial separation coefficient, $s$, which scales the normalized displacement vector $\overrightarrow{d}_{SV}$ between the center-of-masses of the spider and target particle;
more specifically, the oriented spider is initialized at position $\overrightarrow{v} + s\overrightarrow{d}_{SV}$ where $\overrightarrow{v}$ represents the center of mass of the target particle.
Most importantly, the shell and the spider interact via a potential that governs the degree to which (i) the target particle is extracted, and (ii) the remaining shell particles stay intact.
This interaction is driven by a Morse potential between the head particle and shell particles with depth $\varepsilon_{H}$, width $\alpha_{H}$, and equilibrium distance $\sigma_{H} = r_{HP} + r_{V}$.
Note that during optimization, we optimize with respect to $log(\varepsilon_{H})$.
Furthermore, the head particle, base particles, and spider legs interact with the shell-comprising particles via soft sphere repulsion with $\varepsilon_{ss} = 10^4$.\\

\textit{Appendix C: Simulation Details.---}Simulations were performed using the rigid body framework implemented in JAX-MD~\cite{schoenholz2020jax}.
We used a Langevin integrator with a damping constant of $\gamma=10.0$, $kT = 1.0$, and $dt = 0.001$.
To initialize the shell-spider complex, all icosahedron and octahedron simulations used an initial separation coefficient of $s = 0.2$ and $s = 0.0$, respectively.\\

\textit{Appendix D: Optimization Details.---}Each optimization consisted of 5000 iterations of gradient descent using the Adam optimizer with a learning rate of $0.01$ on an 80 GB NVIDIA A100.
In each iteration, we performed a gradient update using the average gradient computed with respect to a batch of 10 simulations initialized with random velocities.
Simulations consisted of either 1000 (Figure 2) or 1500 (Figure 3E) time steps.
Plots of loss values (Figure 2) are truncated after convergence and therefore do not represent all 5000 iterations.

\begin{figure}[b]
\begin{center}
\vspace{-10pt}
\centerline{\includegraphics[width=0.9\columnwidth]{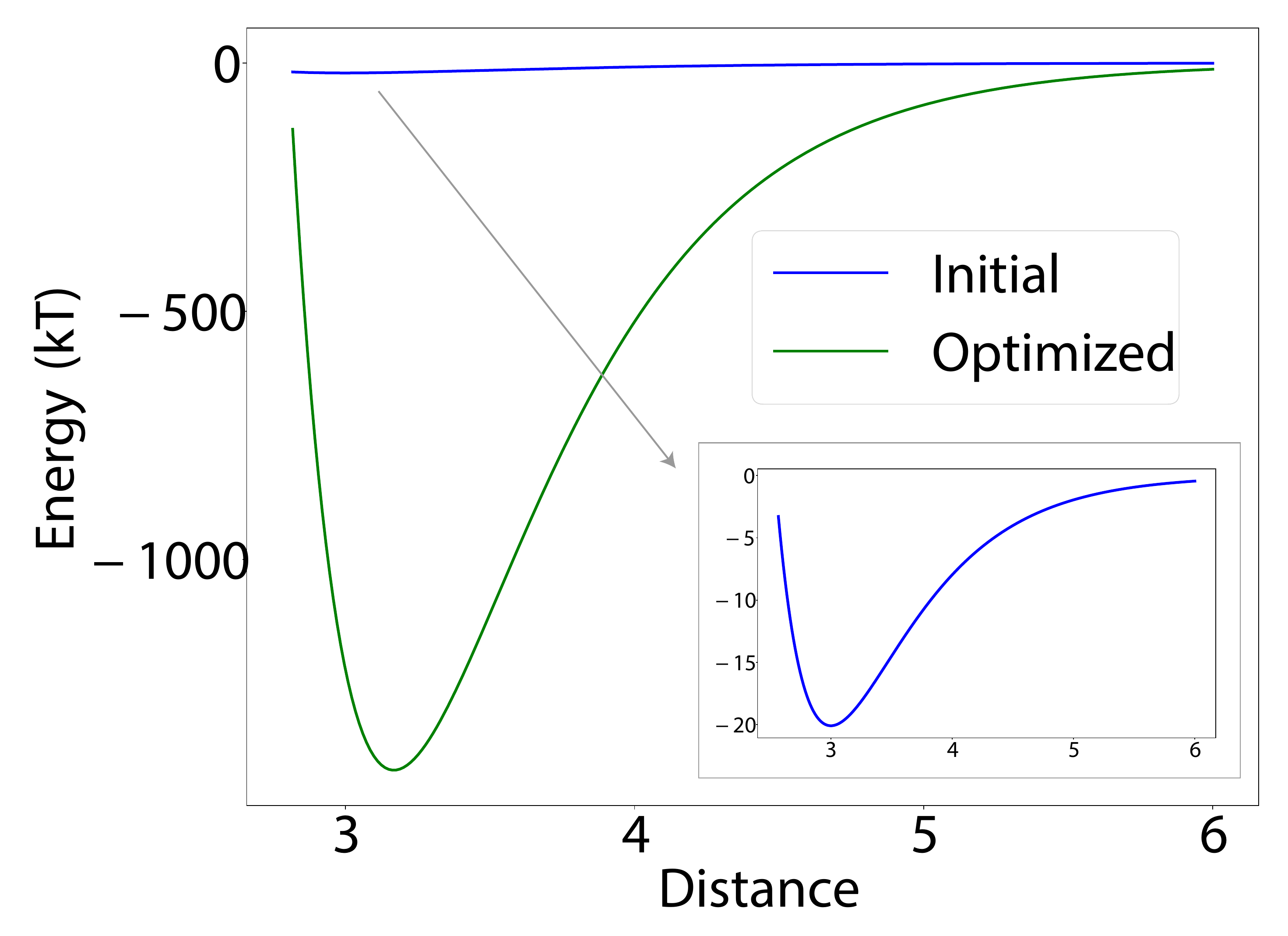}}
\vspace{-10pt}
\caption{
A comparison of the initial and optimized Morse potentials in the weak interaction limit for the simple icosahedron.
Inset: the rescaled initial Morse potential.
}
\label{fig:diffusive-pots}
\end{center}
\end{figure}

\begin{table*}[t]
\resizebox{0.9\linewidth}{!}{
\begin{tabular}{@{}cccccccccc@{}}
\toprule
Initialization & Shell Type & $d$ & $\ell$ & $r_{BP}$ & $r_H$ & $\log (\epsilon_H)$ & $\alpha_H$ & $r_{on}$ & $r_{cut}$ \\ \midrule
\multirow{2}{*}{Weak}   & Icosahedron         & 9.42         & 7.05            & 1.01              & 1.17           & 7.26                         & 1.91                & 9.74              & 10.29              \\
                        & Octahedron          & 8.47         & 7.45            & 1.35              & 0.80           & 8.94                         & 1.70                & 10.51             & 10.38              \\
\multirow{2}{*}{Strong} & Icosahedron         & 9.06         & 7.27            & 1.04              & 0.33           & 10.05                        & 1.97                & 9.83              & 11.79              \\
                        & Octahedron          & 9.89         & 7.47            & 1.65              & 0.52           & 7.67                         & 1.52                & 9.73              & 11.53              \\ \bottomrule
\end{tabular}
}
\caption{
Optimized parameters for all icosahedral optimizations presented in Figure 2 and all octahedral optimizations presented in Figure S\ref{fig:octahedron}.
In all cases, we use the following initial parameter values: $d = 10$, $\ell = 7.07$ (corresponding to a head height of $5.0$), $r_{BP} = 1.0$, $r_H = 1.0$, $r_{on} = 10.0$, and $r_{cut} = 12.0$.
}
\label{tab:opt-params}
\vspace{-10pt}
\end{table*}

For the optimizations shown in Figure 2, we optimize 8 parameters: the (log) attraction strength ($\log (\epsilon_H)$), width ($\alpha_H$), onset ($r_{on}$), and cutoff ($r_{cut}$) of the Morse potential between the head and the shell particles, the radius of the base particles ($r_{BP}$), the diameter of the spider ($d$), the spider leg length ($\ell$), and the radius of the head particle $r_H$.
We impose a minimum value for $r_H$ of $0.1$.
For the free energy diagrams depicted in Figure 3C, we optimize 10 parameters characterizing the extended rigid spider and use the optimized parameters for all free energy calculations.
These 10 parameters are the (log) attraction strength ($\log (\epsilon_A)$), width ($\alpha_A$), onset ($r_{on}$), and cutoff ($r_{cut}$) of the Morse potential between the attractive sites and the shell particles, the radius of the base particles ($r_{BP}$), the diameter of the spider ($d$), the spider leg length ($\ell$), the radius of the head particle ($r_H$), the radius of the attractive site particles ($r_A$), and the relative position of the attractive site on the spider leg ($p_A$).
For the optimizations presented in Figure 3E, we fix the parameters used in Figure 3C and optimize over 5 spring constants corresponding to the bond strength between nearest neighbors and an additional 5 spring constants corresponding to the bond strength between next-nearest neighbors.
Note that only bonds connecting nearest neighbors interact with shell particles via soft-sphere repulsion.\\

\textit{Appendix E: Optimized Parameters--}
Table \ref{tab:opt-params} contains all optimized parameters for optimizations presented in Figure 2 and Figure S\ref{fig:octahedron}.
In the weak and strong interaction limits for the icosahedron (Figure 2A), we initialize $\alpha_H = 1.5$ in both cases and $\log (\epsilon_H) = 3.0$ and $\log (\epsilon_H) = 10.5$, respectively.
In the weak and interaction limits for the octahedron (Figure S\ref{fig:octahedron}), we initialize $\alpha_H = 1.5$ and $\log (\epsilon_H) = 3.5$, and $\alpha_H = 1.0$ and $\log (\epsilon_H) = 8.0$, respectively.
Note that in the weak interaction limit for the icosahedron, $r_{cut} < r_{on}$ in the optimized parameter set.
If desired, one could trivially impose $r_{on} < r_{cut}$ as a constraint during optimization.

While only single optimizations are presented in Figure 2, we additionally perform ensembles of optimizations over different initial seeds for the random number generator and over different random perturbations to the initial $log(\epsilon_{H})$ parameter in both the weak and strong initialization regimes.
Despite the wide variance in initial parameters, we observe low variance in the converged parameter values in all cases, particularly in the strong interaction limit.
We observe the highest variance for $\log (\epsilon_H )$ in the weak interaction limit.

To obtain parameters for the free energy calculations in Figure 3C, we first optimize the parameters of the rigid variant of the modified spider.
This optimization yields the following: $\log (\epsilon_A) = 4.29$, $\alpha_A = 1.42$, $r_{on} = 10.0$, $r_{cut} = 12.0$, $p_A = 0.36$, $r_A = 1.48$, $r_{BP} = 1.50$, $d = 9.28$, $\ell = 10.44$, $r_H = 1.0$.
These parameters are used for all three free energy calculations in Figure 3C to isolate the effect of configurational entropy. Note that in Figure 3C we do not require the spiders extract the particle without disturbing the remaining shell.
In the optimization depicted in Figure 3E, we fix these optimized values and optimize over only the spring constants defining flexible bonds between pairs of nearest and next-nearest neighbors.
All log spring constants are initialized to $2.0$ and the optimized log spring constants are as follows: $k_{12} = -0.73$, $k_{23} = 2.40$, $k_{34} = -0.97$, $k_{45} = -1.12$, $k_{51} = 2.08$, $k_{13} = -0.78$, $k_{24} = -1.03$, $k_{35} = 3.19$, $k_{41} = -2.44$, and $k_{52} = 2.78$ where $k_{ij}$ is the log spring constant of the bond connecting base particles $i$ and $j$.\\

\textit{Appendix F: Energy Functions--}
We use the following functional forms for the Morse potential, soft-sphere repulsion, and harmonic spring, respectively:
\begin{align}
    U_{morse}(r ; \sigma, \epsilon, \alpha) &= \epsilon \left(1 - \exp \left(-\alpha \left(r - \sigma\right)\right)\right)^2 - \epsilon \\
    U_{ss}(r; \sigma, \epsilon) &= \frac{\epsilon}{2} \cdot (1 - \frac{r}{\sigma})^2 \\
    U_{spring}(r ; r_0, k) &= \frac{k}{2} (r - r_0)^2
\end{align}

\textit{Appendix G: Free Energy Calculations--}
All free energy diagrams are calculated via the Weighted Histogram Analysis Method (WHAM).
WHAM involves performing a series of simulations in which each simulation is biased via a harmonic spring centered at given value of the order parameter and the resulting bins of sampled order parameter values are used to compute a free energy diagram via an unbiasing procedure~\cite{kumar1992weighted}.
To resolve the sharp peak in the Morse potential, more bins with a higher spring constant are used for lower distances.
For each free energy diagram, we perform 250 biased simulations at evenly spaced values between $7.0$ and $10.0$ with a spring constant of $10^4$ and an additional 250 biased simulations at evenly spaced values between $10.0$ and $20.0$ with a spring constant of $500$.
For each value of the order parameter, we obtain an initial state via minimizing for $25,000$ steps a spider-particle configuration in which the particle is placed at the corresponding distance from the spider head.
We then sample $40,000$ states every $100$ timesteps.
Unbiasing is performed via the implementation of Grossfield with a tolerance of $10^{-3}$~\cite{Grossfield}.\\

\textit{Appendix H: Scaled Loss Term--}
When optimizing the spring constants of the base bonds for the flexible spider (Figure 3E), we use a modified extraction loss term such that the optimization algorithm has more freedom to promote release in favor of extraction.
Specifically, we apply an inverted ReLU to the original $\mathcal{L}_{extract}$ so that the loss term changes minimally beyond a specified maximum value:
\begin{equation*}
\mathcal{L}_{extract}'(V) = \begin{cases}
    x & \text{if $x \geq x^*$}\\
    x^* + mx & \text{otherwise}
\end{cases}
\end{equation*}
where $x^* = \mathcal{L}_{extract}(V)$ is the original extraction loss and $x^*$ and $m$ are hyperparameters.
In practice, we set $x^* = -7.0$ and $m = 10^{-4}$.
The total loss term for the optimization in Figure 3E therefore is:
\begin{align*}
    \mathcal{L} = \mathcal{L}_{remain}(V_{init}, A_{init}) &+ \mathcal{L}_{extract}'(V_{1000}) \\
    &+ \beta \cdot \mathcal{L}_{release}(\overrightarrow{v}_{1500}, A_{1500})
\end{align*}
where $\beta$ is a hyperparameter, $V_{init}$ and $A_{init}$ are the initialized shell particles and attractive site positions, $V_{1000}$ is the set of shell-comprising particles at the $1000^{th}$ timestep, and $\overrightarrow{v}_{1500}$ and $A_{1500}$ are the target particle and attractive sites at the $1500^{th}$ timestep, respectively.
In practice, we set $\beta = 0.1$.

\bibliography{z_appendix}